
\magnification=1200
\baselineskip=12pt
\overfullrule=0pt

\rightline{UR-1361\ \ \ \ \ \ }
\rightline{ER40685-811}

\bigskip

\baselineskip=18pt

\centerline{\bf ON THE WARD IDENTITIES AT FINITE TEMPERATURE}

\medskip

\centerline{Ashok Das}

\centerline{and}

\centerline{Marcelo Hott$^\dagger$}

\centerline{Department of Physics and Astronomy}

\centerline{University of Rochester}

\centerline{Rochester, NY 14627}

\vskip 1 truein

\centerline{\underbar{Abstract}}

\medskip

We show both in 1+1 and 3+1 dimensions that, contrary to the recent
suggestions, the contribution of the fermion loop to the polarization
tensor is manifestly transverse at finite temperature.  Some subtleties
associated with the Ward identities at finite temperature are also pointed
out.

\vskip 2.5in

\noindent $^\dagger$On leave
 of absence from UNESP - Campus de
Guaratinguet\'a, P.O. Box 205, CEP : 12.500, Guaratinguet\'a, S.P., Brazil

\vfill\eject

\noindent {\bf I. \underbar{Introduction}:}

\medskip

Studies in finite temperture field theory have been of much interest in
recent years from various points of view [1-3].  There are three distinct
formalisms for carrying out calculations at finite temperature, namely, the
imaginary time formalism [4], thermofield dynamics [5] and the closed time
path formalism [6].  While the nature of a  finite temperature calculation,
in each of these formalisms, is completely parallel to the zero temperature
calculation, some of the resulting properties at finite temperature are
quite distinct.  For example, the analytic behavior of Feynman amplitudes
at finite temperature is noticeably different from those at zero
temperature [7-9], the infrared behavior at finite temperature is markedly
more singular [10] and so on.  As more and more studies are carried out, we
will no doubt have a better understanding of the structure of quantum field
theory at finite temperature.

It is commonly believed that gauge invariance is not affected by
temperature.  The recent suggestion that Ward identities may be violated at
finite temperature [11] is, therefore, quite surprising.  More
specifically, it was argued that the contribution of the fermion loop, at
finite temperature, to the polarization tensor is not transverse with
$\epsilon$-regularization [8].  We note that the physical component of the
fermion propagator at finite temperature has the form

\medskip

$$\eqalign{S(p) &= S^{(0)} (p) + S^{(\beta)} (p)\cr
\noalign{\vskip 5pt}%
&= i(\rlap \slash{\rm p} + m) \bigg( {1 \over p^2 - m^2 + i \epsilon} +
2i \pi n_F (p) \delta (p^2 -m^2)\bigg)\cr}\eqno(1)$$

\medskip

\noindent where

\medskip

$$n_F (p) = {1\over e^{\beta | p_0 |} +1} \eqno(2)$$

\medskip

\noindent represents the fermion distribution function.  (One can write this
more covariantly by introducing the velocity of the heat bath, but we will
ignore it in this initial discussion.)  The temperature dependent
contribution of the fermion loop

\vfill\eject

\ \ \ \

\vskip 1 truein

\noindent to the polarization tensor is, then, given by ($D$ is the number
of space-time dimensions)
$$\eqalign{-i \Pi^{(\beta)}_{\mu \nu} (p) &= e^2 \int {d^D k \over
(2 \pi)^D}\ {\rm Tr} \bigg[ \gamma_\mu S^{(0)} (k + p) \gamma_\nu
S^{(\beta)} (k) + \gamma_\mu S^{(\beta)} (k+p) \gamma_\nu
S^{(0)} (k)\cr
\noalign{\vskip 4pt}%
& \qquad\qquad\qquad\qquad\qquad
+ \gamma_\mu S^{(\beta)} (k+p) \gamma_\nu S^{(\beta)} (k) \bigg]\cr
\noalign{\vskip 4pt}%
&=-D e^2 \int {d^D k \over (2 \pi)^D} \ \big[ (k+p)_\mu k_\nu + (k+p)_\nu
k_\mu - \eta_{\mu \nu} (k \cdot (k+p) -m^2) \big]\cr
\noalign{\vskip 4pt}%
&\ \quad \times \ \bigg[ {2 i \pi n_F (k) \over (k+p)^2 -m^2
+ i \epsilon}\ \delta (k^2 -m^2) +
{2 i \pi n_F (k+p) \over k^2 -m^2 + i \epsilon}\ \delta ((k+p)^2 -m^2)\cr
\noalign{\vskip 4pt}%
&\qquad\qquad\qquad - 4 \pi^2 n_F (k)
n_F(k+p) \delta (k^2-m^2) \delta ((k+p)^2 - m^2) \bigg]\cr}\eqno(3)$$
The three terms in $p^\mu \Pi^{(\beta)}_{\mu \nu} (p)$ can now be easily
seen to vanish individually due to the $\delta$-function constraints
and the symmetry of the respective integrands.  It is, therefore, even more
surprising that an explicit evaluation of the polarization tensor will be
nontransverse considering that temperature dependent contributions are
finite.

In this paper, we report on a systematic study of the transverse nature of
the fermion loop contribution to the polarization tensor at finite
temperature.  In sec. II, we recapitulate various known facts about
distinct tensor structures at finite temperature.  In sec. III, we show
that in 1+1 dimensions, the fermion bubble has a transverse tensor
structure which factors out of the integral.  In sec. IV, we show that
the integrand for the fermion bubble is manifestly transverse even in 3+1
dimensions.  In sec. V, we point out some subtleties associated with the
Ward identities at finite temperature and present our conclusions in sec.
VI.

\medskip

\noindent {\bf II. \underbar{Tensor Structures at Finite $T$}:}

\medskip

In this brief section, we collect all the known facts [12] about various
tensor structures at finite temperature for completeness as well as for
later use.  We note that a covariant description, at finite temperature,
involves the velocity of the heat bath, $u^\mu$, in addition to the
available four momenta.  Any vector can, of course, be decomposed into
components parallel and perpendicular to $u^\mu$ as $(u^2 = u^\mu
u_\mu = 1)$
$$\eqalign{A^\mu_\parallel &= (A \cdot u)u^\mu\cr
A^\mu_\perp &= \widetilde A^\mu = A^\mu -(A \cdot u) u^\mu\cr}\eqno(4)$$
In particular, we note that for a given momentum $p^\mu$, if we define
$$\omega = (p \cdot u) \eqno(5)$$
Then, the component of $p^\mu$ orthogonal to $u^\mu$ will be given by
$$\widetilde p^\mu = p^\mu - \omega u^\mu \eqno(6)$$
It follows from this that
$$\widetilde p^2 = \widetilde p^\mu \widetilde p_\mu = p^2 - \omega^2
\eqno(7)$$

\noindent From Eq. (4),
 it is clear that the projection operator onto the space of
vectors transverse to $u^\mu$ is given by
$$\widetilde \eta^{\mu \nu} = \eta^{\mu \nu} - u^\mu u^\nu \eqno(8)$$
Here $\eta^{\mu \nu}$ is the metric and in our convention has the signature
$(+,-,-,-)$.

One can, of course, decompose any vector with respect to a given momentum
$p^\mu$ as well.  Thus, the component of a vector $A^\mu$, transverse to
$p^\mu$ is given by
$$\overline A^\mu = A^\mu - {1 \over p^2} \ (p \cdot A) p^\mu \eqno(9)$$
In particular, this gives
$$\overline u^\mu = u^\mu - {\omega \over p^2}\ p^\mu \eqno(10)$$
which is orthogonal to $p^\mu$.  Note, from Eq. (9), that the projection
operator onto the space of vectors orthogonal to $p^\mu$ is given by
$$\overline \eta^{\mu \nu} = \eta^{\mu \nu} - {p^\mu p^\nu \over
p^2} \eqno(11)$$

Given these, it is easy to check that at finite temperature, we can
construct only three second rank, symmetric tensors from $p^\mu$ and
$u^\mu$ which are orthogonal to $p^\mu$, namely,
$$\eqalign{A^{\mu \nu} &= \eta^{\mu \nu} - {p^\mu p^\nu \over p^2}\cr
B^{\mu \nu} &= \widetilde \eta^{\mu \nu} - {\widetilde p^\mu
\widetilde p^{\, \nu} \over
\widetilde p^2} \cr
C^{\mu \nu} &= {p^2 \over \widetilde p^{\, 2}} \ \overline u^\mu \overline
u^\nu \cr}\eqno(12)$$
Each of these structures is normalized so that they act like projection
operators.  In fact, they satisfy
$$\eqalign{&A^{\mu \nu} A_{\nu \lambda} = A^\mu_{\ \lambda} \cr
&A^{\mu \nu}B_{\nu \lambda} = B^\mu_{\ \lambda}\cr}
\qquad\qquad \eqalign{&B^{\mu \nu} B_{\nu \lambda} = B^\mu_{\ \lambda}\cr
&A^{\mu \nu} C_{\nu \lambda} = C^\mu_{\ \lambda}\cr}
\qquad\qquad \eqalign{&C^{\mu \nu} C_{\nu \lambda} = C^\mu_{\ \lambda}\cr
&B^{\mu \nu} C_{\nu \lambda} = 0\cr}\eqno(13)$$
It is also easy to check that the three structures are not independent.  In
fact,
$$A^{\mu \nu} = B^{\mu \nu} + C^{\mu \nu} \eqno(14)$$
and since $B^{\mu \nu}$ and $C^{\mu \nu}$ are orthogonal to each other, it
is more useful to treat them as independent.  Given this, then, one can
write down the most general transverse polarization tensor at finite
temperature to have the form
$$\eqalign{\Pi^{(\beta)}_{\mu \nu} (p) &= B_{\mu \nu} \Pi^{(\beta)}_T (p) +
C_{\mu \nu} \Pi^{(\beta)}_L (p)\cr
\noalign{\vskip 4pt}%
&= \bigg( \widetilde \eta_{\mu \nu} - {\widetilde p_\mu
\widetilde p_\nu \over
\widetilde p^{\, 2}} \bigg) \Pi^{(\beta)}_T (p) + {p^2 \over \widetilde
 p^{\, 2}}
\ \overline u_\mu \overline u_\nu
\Pi_L^{(\beta)} (p)\cr}\eqno(15)$$
At zero temperature, $\Pi_T (p) = \Pi_L (p)$ and with the identity (14),
 we see
that the polarization tensor reduces to the usual structure.  It is also
 easy to
check from Eq. (15) that
$$\eqalign{\Pi^{(\beta)}_L &= {p^2 \over \widetilde p^{\, 2}} \ u^\mu u^\nu
\Pi^{(\beta)}_{\mu \nu}\cr
(D-2)\Pi^{(\beta)}_T &= \eta^{\mu \nu} \Pi_{\mu \nu} - \Pi^{(\beta)}_L
\cr}\eqno(16)$$
where $D$ is the number of space-time dimensions.  We note that Eq.
(16) does not yield any information on $\Pi^{(\beta)}_T$ when $D=2$.

\medskip

\vfill\eject

\noindent {\bf III. \underbar{Polariation Tensor in 1+1 Dimensions}:}

\medskip

In this section, we study the contribution of the fermion bubble to the
polarization tensor at finite temperature in 1+1 dimensions.  From Eq. (3),
we note that
$$\eqalign{-i \Pi^{(\beta)}_{\mu \nu} (p) = &- {i e^2 \over \pi}
\int d^2k \big[ (k+p)_\mu k_\nu + ( k+p)_\nu
k_\mu - \eta_{\mu \nu} (k \cdot (k+p) - m^2 )\big]\cr
&\times\ \bigg[ {n_F (k) \over (k+p)^2 -m^2 + i \epsilon}\
\delta (k^2 -m^2) +
 {n_F (k+p) \over k^2 -m^2 + i \epsilon}\ \delta ((k+p)^2 -m^2)\cr
&\qquad  + 2i \pi n_F (k) n_F (k+p)
\delta (k^2 - m^2) \delta ((k+p)^2
-m^2 )\bigg]\cr}\eqno(17)$$
where the covariant fermion distribution function is given by
$$n_F (k) = {1 \over e^{\beta |u\cdot k|} +1} \eqno(18)$$
The simplicity of 1+1 dimensions allows one to factor the tensor structure
 out of the integral.  To see this, let us define
$$\eqalign{\Omega &= k \cdot u \qquad\qquad \omega = p \cdot u\cr
k^\mu &= \Omega u^\mu - \epsilon^{\mu \nu} u_\nu k^\prime\cr
p^\mu &= \omega u^\mu - \epsilon^{\mu \nu} u_\nu p^\prime\cr}\eqno(19)$$
In our notation $\epsilon^{01} = 1$ and $k^{\prime 2} = - \widetilde k^2$.
Substituting these new variables into Eq. (17) (The Jacobian can be easily
checked to be unity.) and with some straightforward algebraic manipulations
we find that the $\delta$-function constraints as well as the symmetry
properties of the integrals give
$$-i \Pi^{(\beta)}_{\mu \nu} (p) = -e^2 \overline u_\mu
\overline u_\nu (2I + \widetilde I) \eqno(20)$$
where, in 1+1 dimensions, we have
$$\overline u_\mu = u_\mu - {\omega \over p^\prime}\ \epsilon_{\mu \nu}
u^\nu \eqno(21)$$
and
$$\eqalign{I &= {i p^\prime \over \omega} \int d\Omega dk^\prime {k^\prime
(\omega + \Omega) + \Omega (k^\prime + p^\prime) \over \omega (\omega +
2 \Omega)-p^\prime (p^\prime + 2k^\prime) + i \epsilon}\
{1 \over e^{\beta |\Omega |} + 1} \ \delta \big( \Omega^2 -
k^{\prime 2} - m^2)\cr
\widetilde I &= -{2 p^\prime \over \omega} \int d\Omega dk^\prime
(k^\prime (\omega + \Omega) + \Omega (k^\prime + p^\prime))
{1 \over e^{\beta |\Omega |}+1}\ {1 \over e^{\beta | \omega + \Omega|} +1}
\ \delta (\Omega^2 - k^{\prime 2} - m^2) \cr
&\qquad\qquad\qquad \times \delta ((\Omega + \omega)^2
-(k^\prime + p^\prime )^2 - m^2)\cr}\eqno(22)$$

\line{\indent We  \hfill
 note  \hfill that  \hfill the  \hfill temperature \hfill
 dependent  \hfill part  \hfill of  \hfill the  \hfill
 polarization  \hfill tensor  \hfill is  \hfill
explicitly}

\line{transverse \hfill even \hfill
  before \hfill evaluating \hfill the \hfill integrals \hfill
 and, \hfill  therefore, \hfill
any \hfill prescription}

\noindent  ($\epsilon$-regularization, modified Feynman
parameterization etc.) for evaluating the integral cannot change this.
Comparing with Eq. (15), we note that in this case,
$$\Pi^{(\beta)}_T (p) = 0 \eqno(23)$$
If $m=0$, then one can evaluate the integrals in Eq. (22) in a
straightforward manner and show that
$$\eqalign{&{\rm Im}\ I = 0\cr
&{\rm Re}\ (2I + \widetilde I) = 0 \cr}\eqno(24)$$
so that for $m=0$
$$\eqalign{&\Pi^{(\beta)}_L = 0 \cr
&\Pi^{(\beta)}_{\mu \nu} = 0 \cr} \eqno(25)$$
This is indeed consistent with what one knows about the Schwinger model [13
-15] where the temperature dependent corrections to the photon mass and the
chiral anomaly vanish.  When $m \not= 0$, however, $\Pi^{(\beta)}_L$ does
not vanish and gives the only temperature dependent correction to
$\Pi^{(\beta)}_{\mu \nu}$.  In this case it is easy to see that
Re $\Pi^{(\beta)}_L (p)$ is nonanalytic at $p^\mu =0$ much like the scalar
theories.

\medskip

\noindent {\bf IV. \underbar{Polarization Tensor in 3+1 Dimensions}:}

\medskip

The structure of the polarization tensor in 3+1 dimensions is still the
same as in 1+1 dimensions, namely,
$$\eqalign{-i \Pi^{(\beta)}_{\mu \nu} = -{ie^2  \over 2\pi^3} &\int d^4k
\big[
(k+p)_\mu k_\nu + (k+p)_\nu k_\mu
 - \eta_{\mu \nu} (k \cdot (k+p) -m^2) \big]\cr
\times\ &\bigg[ {n_F (k) \over (k+p)^2 -m^2 + i \epsilon}\ \delta(
k^2 -m^2) + {n_F
(k +p) \over k^2 -m^2 + i \epsilon}\ \delta ((k+p)^2 -m^2)\cr
&\qquad + 2i
\pi n_F (k) n_F (k+p) \delta (k^2 -m^2)\delta ((k+p)^2 -m^2 )\bigg]\cr}
\eqno(26)$$
However, in 3+1 dimensions, it is not possible to factor the tensor
structure out of the integral [16].

To proceed in this case, therefore, we decompose vectors into components
parallel and perpendicular to $p^\mu$.  Thus, let (This is different from
 $\Omega$ of the last section.)
$$\eqalign{\Omega &= p \cdot k \qquad\qquad \omega = u \cdot p\cr
k^\mu &= \Omega \  {p^\mu \over p^2} + \overline k^\mu\cr
u^\mu &= {\omega \over p^2}\ p^\mu + \overline u^\mu\cr}
\eqno(27)$$
It is clear that $\overline k^\mu$ is manifestly transverse to $p^\mu$.
Substituting this back into Eq. (26) and using the $\delta$-function
constraint as well as the symmetry properties of the integrals, we obtain
$$\eqalign{-i \Pi^{(\beta)}_{\mu \nu} (p) = &- {i e^2 \over \pi^3} \int d
\Omega d^4 \overline k\  \delta (p \cdot \overline k) \bigg[ (
- \eta_{\mu \nu} + {p_\mu p_\nu \over p^2}) {\Omega \over 2} +
\overline k_\mu \overline k_\nu \bigg]\cr
&\times\ \bigg[ {2 \over p^2 +2\Omega + i \epsilon} \
{1 \over e^{\beta |{\omega
\Omega \over p^2} + u \cdot \overline k|} +1} \ \delta
\bigg( {\Omega^2 \over p^2} + \overline k^2 - m^2 \bigg)\cr
&+ 2i\pi {1 \over e^{\beta |{\omega \Omega \over p^2} + u \cdot \overline k|}
+1} \ {1 \over e^{\beta|\omega (1+ {\Omega \over p^2})+u\cdot \overline k|}+1}
\cr
&\times\ \delta \bigg( {\Omega^2 \over p^2} + \overline k^2 -m^2 \bigg) \delta
(p^2 + 2 \Omega ) \bigg]\cr}\eqno(28)$$
We note that the first tensor structure is nothing other than $A_{\mu \nu}
= B_{\mu \nu} + C_{\mu \nu}$ and can be factored out of the integral.
While the second structure cannot be factored out of the integral, we note
that it is manifestly orthogonal to $p^\mu$.  Thus, independent of the
prescription used to evaluate this, it can only lead to a linear
combination of $B_{\mu \nu}$ and $C_{\mu \nu}$.  Thus, we see that the
temperature dependent contribution of the fermion bubble to the
polarization tensor continues to be manifestly transverse even in 3+1
dimensions.  Contrary to the 1+1 dimensions, however, in the present case
both $\Pi^{(\beta)}_T$ and $\Pi^{(\beta)}_L$ are a priori nonvanishing.

\medskip

\noindent {\bf V. \underbar{Observations  on Ward Identities}:}

\medskip

At zero temperature, the Ward identities of QED can be written in several
equivalent ways.  For example, if
$$S^{(0)} (p) = {i \over \rlap \slash{\rm p} - m + i \epsilon} \eqno(29)$$
denotes the tree level fermion propagator at zero temperature, then
$${\partial S^{(0)} (p) \over \partial p^\mu} = - {1 \over e}\ S^{(0)} (p)
\Gamma^{(0)}_\mu S^{(0)} (p)\eqno(30)$$
where
$$\Gamma^{(0)}_\mu = -ie \gamma_\mu \eqno(31)$$
is the vertex at the tree level.  The identity in Eq. (30) can be shown to
hold order by order so that to all orders one can write
$${\partial S(p) \over \partial p^\mu}= - {1 \over e}\  S(p) \Gamma_\mu
S(p) \eqno(32)$$
where $S(p)$ and $\Gamma_\mu$ represent the complete fermion propagator and
the vertex of the theory.  This is one form of the Ward identity and, in
fact, is the form in which it was originally derived [17].  The integrated
form of Eq. (32) which can also be derived from the BRS invariance [18] of
the theory is given by
$$(p^\prime -p)^\mu S(p^\prime) \Gamma_\mu S(p) = e(S(p) - S(p^\prime))
\eqno(33)$$
At zero temperature both the forms of the Ward identity are equivalent.

At finite temperature, however, the propagators have a $2 \times 2$ matrix
structure [5-6].  Thus, for example, the fermion propagator can be written
as
$$S(p) = \pmatrix{S_{++} (p) & S_{+-}(p)\cr
\noalign{\vskip 4pt}%
S_{-+} (p) &S_{--}(p)\cr} \eqno(34)$$
where at tree level, the components of the matrix have the form
$$\eqalign{S_{++}(p) &= i(
\rlap \slash{\rm p} + m) \bigg( {1 \over p^2 -m^2 + i \epsilon} + 2i \pi
n_F (p) \delta (p^2 -m^2 )\bigg)\cr
\noalign{\vskip 4pt}%
S_{+-}(p) &= i(\rlap \slash{\rm p} +m) (-2i\pi \theta(-p^0) + 2i \pi
n_F (p) )\delta (p^2 -m^2)\cr
\noalign{\vskip 4pt}%
S_{-+} (p) &= i(\rlap \slash{\rm p} +m) (-2i \pi \theta (p^0) + 2i\pi
n_F (p)) \delta (p^2 -m^2)\cr
\noalign{\vskip 4pt}%
S_{--}(p) &= i(
\rlap \slash{\rm p} + m) \bigg(- {1 \over p^2 -m^2 + i \epsilon} + 2i \pi
n_F (p) \delta (p^2 -m^2 )\bigg)\cr}\eqno(35)$$
The vertex at finite temperature also has a $2 \times 2$ matrix structure
which at the tree level is given by
$$\Gamma_\mu = -ie \pmatrix{\gamma_\mu &0\cr
\noalign{\vskip 6pt}%
0&-\gamma_\mu\cr}\eqno(36)$$

One can now check explicitly and show that the identity in Eq. (33)
continues to hold even at finite temperature be it in terms of matrices.
The identity in Eq. (30), however, modifies at finite temperature (as can
be checked explicitly from the tree level functions defined in Eqs. (35)
and (36)) to
$${\partial S(p) \over \partial p^\mu} = - {1 \over e}\ S(p)
\Gamma_\mu S(p) - 2\pi {\partial n_F (p) \over
\partial p^\mu} \ (\rlap \slash{\rm p} +m) \delta (p^2 -m^2)
\pmatrix{1&1\cr
1&1\cr} \eqno(37)$$
The additional term in Eq. (37), however, can be seen to vanish when
multiplied by $(\rlap \slash{\rm p} -m)$ or $S^{-1}(p)$.  Let us also note
that the retarded and the advanced propagators defined as
$$\eqalign{S_R (p) &= S_{++} (p) -S_{+-} (p)\cr
S_A (p) &= S_{++} (p) - S_{-+} (p)\cr}\eqno(38)$$
on the other hand, can be seen to satisfy
$$\eqalign{{\partial S_R (p) \over \partial p^\mu} &= -S_R (p) \gamma_\mu S
_R (p)\cr
\noalign{\vskip 4pt}%
{\partial S_A (p) \over \partial p^\mu} &= -S_A (p) \gamma_\mu S_A (p)\cr}
\eqno(39)$$
without the extra term in Eq. (37).  Thus, one should be careful at finite
temperature in the use of the form of  Ward identity.

\medskip

\noindent {\bf VI. \underbar{Conclusion}:}

\medskip

In this paper, we have systematically studied the transverse nature of the
fermion bubble contribution to the polarization tensor at finite
temperature.  Contrary to the recent suggestions, we have shown that the
polarization tensor is manifestly transverse.  In 1+1 dimensions the
transverse tensor structure factors out of the integral and yields a
nonvanishing contribution only to $\Pi^{(\beta)}_L$ when $m\not=0$.  In 3+1
dimensions, the tensor structure does not factor out of the integral, but
is manifestly transverse.  We have also pointed out some subtleties
associated with the Ward identities at finite temperature.

One of us (A.D.) would like to thank Prof. E. Witten and the Institute for
Advanced Studies for hospitality where part of this work was done.  This
work was supported in part by the U.S. Department of Energy Grant No.
DE-FG-02-91ER40685.    M.H. would like to thank the
Funda\c c\~ao de Amparo a Pesquisa do Estado de S\~ao
 Paulo for the financial
support.

\vfill\eject

\noindent {\bf \underbar{References}:}

\item{1.} For recent reviews see N.P. Landsman and Ch. G. Van Weert, Phys.
Rep. {\bf 145} (1987) 141; K.C. Chou, Z.B. Su, B.C. Hao and L. Yu, Phys.
Rep. {\bf 118} (1985) 1; E. Calzetta, B.L. Hu and J.P. Paz,
 \lq\lq Statistical and
Kinetic Field Theory", to be published.

\item{2.} E. Braaten and R. Pisarski, Nucl. Phys. {\bf B337} (1990) 569;
E. Braaten and R. Pisarski, Phys. Rev. {\bf D45} (1992) R1827; J. Frenkel
and J.C. Taylor, Nucl. Phys. {\bf B334} (1990) 199; J.C. Taylor and S.M.
Wong, Nucl. Phys. {\bf B346} (1990) 115.

\item{3.} A. Anselm, Phys. Lett. {\bf B217} (1989) 169; J.D. Bjorken, Int.
J. Mod. Phys. {\bf A7} (1992) 4189; K.L. Kowalsky and C.C. Taylor, Case
Western Reserve University preprint 92-6; K. Rajagopal and F. Wilczek,
Nucl. Phys. {\bf B404} (1993) 577; P.F. Bedaque and A. Das, Mod. Phys. Lett.
{\bf A8} (1993) 3151.

\item{4.} For an introduction see A.L. Fetter and J.D. Walecka,
\lq\lq Quantum Theory of Many Particle Systems", McGraw-Hill (1971); A.A.
Abrikosov, L.P. Gorkov and I.E. Dzyaloshinski, \lq\lq Methods of Quantum
Field Theory in Statistical Physics", Dover (1963).

\item{5.} H. Umezawa, H. Matsumoto and M. Tachiki, \lq\lq Thermo Field
Dynamics and Condensed States", North Holland (1982).

\item{6.} J. Schwinger, J. Math. Phys. {\bf 2} (1961) 407; J. Schwinger,
Lecture Notes of Brandeis Summer Institute in Theoretical Physics (1960);
 L.V. Keldysh, Sov. Phys. JETP {\bf 20} (1965) 1018.

\item{7.} For an excellent review of the problem and references see P.S.
Gribosky and B.R. Holstein, Z. Phys. {\bf C47} (1990) 205.

\item{8.} P.F. Bedaque and A. Das, Phy. Rev. {\bf D45} (1992) 2906.

\item{9.} H.A. Weldon, Phys. Rev. {\bf D47} (1993) 594; P.F. Bedaque and A.
Das, Phys. Rev. {\bf D47} (1993) 601.

\item{10.} For a review and references see H. Weldon, West Virginia
University preprint hep-ph/9311245.

\item{11.} N.R. Pantoja, Universidad de los Andes preprint (1994).

\item{12.} We follow here mainly the discussion in H.A. Weldon, Phys. Rev.
{\bf D26} (1982) 1394.

\item{13.} J. Schwinger, Phys. Rev. {\bf 128} (1962) 2425.

\item{14.} L. Dolan and R. Jackiw, Phys. Rev. {\bf D9} (1974) 3320.

\item{15.} A. Das and A. Karev, Phys. Rev. {\bf D36} (1987) 623.

\item{16.} The polarization tensor has also been evaluated in ref. 12 and
 in O.K. Kalashinov and
V.V. Klimov, Sov. J. Nucl. Phys. {\bf 31} (1980) 699.

\item{17.} J.C. Ward, Phys. Rev. {\bf 77} (1950) 293; ibid {\bf 78} (1950)
182; Y. Takahashi, Nuovo Cimento {\bf 6} (1957) 370.

\item{18.} C. Becchi, A. Rouet and R. Stora, Comm. Math. Phys.
{\bf 52} (1975) 55.

\end